\documentclass[letterpaper,twocolumn,american,prb,superscriptaddress,showpacs]{revtex4}
\usepackage{graphicx}
\usepackage{amssymb}

\makeatletter

\newcommand{\noun}[1]{\textsc{#1}}

\usepackage{graphicx}

\usepackage{babel}

\usepackage{babel}
\makeatother
\begin{document}

\title{Length-dependent conductance and thermopower in single-molecule junctions
of dithiolated oligophenylene derivatives}

\author{F.~Pauly}

\email{fabian.pauly@kit.edu}

\affiliation{Institut f\"ur Theoretische Festk\"orperphysik and DFG-Center for
Functional Nanostructures, Universit\"at Karlsruhe, 76128 Karlsruhe,
Germany}

\affiliation{Forschungszentrum Karlsruhe, Institut f\"ur Nanotechnologie, 76021
Karlsruhe, Germany }

\author{J.~K.~Viljas}

\affiliation{Institut f\"ur Theoretische Festk\"orperphysik and DFG-Center for
Functional Nanostructures, Universit\"at Karlsruhe, 76128 Karlsruhe,
Germany}

\affiliation{Forschungszentrum Karlsruhe, Institut f\"ur Nanotechnologie, 76021
Karlsruhe, Germany }

\author{J.~C.~Cuevas}

\affiliation{Departamento de F\'isica Te\'orica de la Materia Condensada, Universidad
Aut\'onoma de Madrid, 28049 Madrid, Spain}

\affiliation{Institut f\"ur Theoretische Festk\"orperphysik and DFG-Center for
Functional Nanostructures, Universit\"at Karlsruhe, 76128 Karlsruhe,
Germany}

\affiliation{Forschungszentrum Karlsruhe, Institut f\"ur Nanotechnologie, 76021
Karlsruhe, Germany }

\date{\today{}}

\begin{abstract}
We study theoretically the length dependence of both conductance and
thermopower in metal-molecule-metal junctions made up of dithiolated
oligophenylenes contacted to gold electrodes. We find that while the
conductance decays exponentially with increasing molecular length,
the thermopower increases linearly as suggested by recent experiments.
We also analyze how these transport properties can be tuned with methyl
side groups. Our results can be explained by considering the level
shifts due to their electron-donating character as well as the tilt-angle
dependence of conductance and thermopower. Qualitative features of
the substituent effects in our density-functional calculations are
explained using a tight-binding model. In addition, we observe symmetry-related
even-odd transmission channel degeneracies as a function of molecular
length.
\end{abstract}

\pacs{85.65.+h, 65.80.+n, 73.23.Ad, 73.63.Rt}

\keywords{molecular electronics; single-molecule contact; thermoelectricity;
Seebeck effect; Seebeck coefficient; length dependence; tilt-angle
dependence}

\maketitle

\section{Introduction}

In the field of molecular electronics, research has so far mostly
concentrated on the dc electrical conduction properties of single-molecule
contacts.\cite{Tao06} By now it is known that the charge transport
through organic molecules is typically due to electron tunneling.
This is evidenced in particular by the exponential decay of the conductance
$G$ with increasing molecular length in contacts formed from oligomers
with varying numbers of units.\cite{Wold02,Wakamatsu06,Venkataraman06}
Considering the statistical nature of experiments at the molecular
scale, a conclusive comparison to theory is presently difficult. Nevertheless,
associated decay coefficients appear to be reproduced by theoretical
calculations.\cite{Kaun03,Kondo04,Su05} For a deeper understanding
of molecular-size contacts, it is useful to analyze also other observables
in parallel with the dc conductance. Emerging new lines of research
involve the photoconductance\cite{Viljas07a,Guhr07a,Viljas07b} and
heat transport.\cite{Galperin07,Wang07} In this paper we concentrate
on the thermopower. For metallic atomic contacts, this quantity was
studied experimentally already some years ago,\cite{Ludoph98} but
for molecular contacts only very recently.\cite{Reddy07,Baheti08}

The thermopower $Q$, also known as the Seebeck coefficient, measures
the voltage $\Delta V$ induced over a conducting material at vanishing
steady state electric current $I$, when a small temperature difference
$\Delta T$ is applied: $Q=\left.-\Delta V/\Delta T\right|_{I=0}$.
It is known that in bulk materials the sign of the thermopower is
a hint of the sign of the main charge carriers. If $Q<0$ ($Q>0$),
charge is carried by electron-like (hole-like) quasiparticle excitations,
as in an $n$-doped ($p$-doped) semiconductor.\cite{AshcroftMermin}
Analogously, the thermopower of the molecular junction gives information
about the alignment of the energies of the highest occupied and lowest
unoccupied molecular orbitals (HOMO and LUMO) with respect to the
metal's Fermi energy $E_{F}$.\cite{Paulsson03,Koch04,Segal05} In
the experiment,\cite{Reddy07} $Q$ was measured for gold electrodes
bridged by dithiolated oligophenylene molecules. It was found to be
positive, which indicates that $E_{F}$ lies closer to the HOMO than
to the LUMO. Also, it was observed that $Q$ grows roughly linearly
with the number $N$ of the phenyl rings in the molecule. More recently\cite{Baheti08}
the effects of substituents and varied end groups have been analyzed
for the benzene molecule.

Using transport calculations based on density-functional theory (DFT)
we investigate in this paper the length dependence of the conductance
and the Seebeck coefficient for dithiolated oligophenylenes bonded
to gold contacts. For the molecules studied in Ref.~\onlinecite{Reddy07},
we find that the conductance decays exponentially with increasing
molecular length. Decay coefficients compare reasonably with previous
DFT calculations\cite{Kaun03,Kondo04,Su05} and also with experiments,\cite{Wold02,Wakamatsu06,Venkataraman06}
considering differences in contact configurations, molecular end groups,
and uncertainties with respect to environmental effects. In addition,
the thermopower increases linearly, with a magnitude coinciding with
the measurements.\cite{Reddy07} We also study how the results change
as various numbers of methyl substituents are introduced to the molecules.
The effect of these substituents is twofold: (i) they push the energies
of the $\pi$ electrons up as a result of their electron-donating
behavior\cite{Venkataraman07,Baheti08} and (ii) they increase the
tilt angles between the phenyl rings through steric repulsion. The
latter effect tends to decrease both $G$ and $Q$ due to a reduction
of the degree of $\pi$-electron delocalization, while the former
opposes this tendency by bringing the HOMO closer to $E_{F}$. A simplified
$\pi$-orbital model is used to explain essential features of the
DFT results.

The paper is organized as follows. Sec.~\ref{sec:Methods} explains
details of our DFT-based approach. Then, in Sec.~\ref{sec:Contacts},
we present the molecular contacts, whose charge transport properties
we determine in Sec.~\ref{sec:Density-functional-based-transport}.
In Sec.~\ref{sec:pi-orbital-model} we show, how the DFT results
can be understood in terms of the $\pi$-orbital model, but we discuss
also effects beyond this simplified picture in Sec.~\ref{sec:beyond-pi-model}.
Finally, we end in Sec.~\ref{sec:Discussion-and-conclusions} with
a discussion and conclusions.

\begin{figure}
\begin{centering}\includegraphics[width=1\columnwidth,clip=]{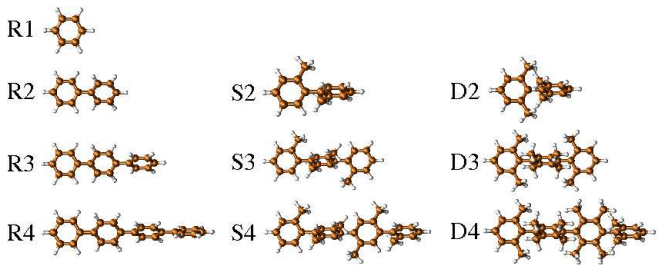}\par\end{centering}

\caption{\label{cap:molecules}(Color online) The molecules studied in this
work. When they are contacted to the gold electrodes, sulfur atoms
replace the terminal hydrogen atoms.}
\end{figure}

\section{Methods\label{sec:Methods}}

The general formulas for the treatment of thermoelectric effects,
based on the Landauer-B\"uttiker formalism, are discussed in detail
in several references.\cite{Sivan86,vanHouten92,Ludoph98,Paulsson03,Segal05,Wang05}
By expanding the expression for the current $I$ to linear order in
$\Delta V$ and $\Delta T$ and considering the cases $\Delta T=0$
and $I=0$, respectively, one arrives at \begin{equation}
G=G_{0}K_{0}(T),\quad Q=-\frac{K_{1}(T)}{eTK_{0}(T)},\label{eq:GandS}\end{equation}
with $K_{0}(T)=\int\tau(E)\left[-\partial f(E,T)/\partial E\right]dE$,
$K_{1}(T)=\int\left(E-\mu\right)\tau(E)\left[-\partial f(E,T)/\partial E\right]dE$,
and $G_{0}=2e^{2}/h$. Here $\tau(E)$ is the transmission function,
$f(E,T)=\left\{ \exp\left[\left(E-\mu\right)/k_{B}T\right]+1\right\} ^{-1}$
the Fermi function, and $\mu$ the chemical potential, $\mu\approx E_{F}$.
At low temperature, the leading-order terms in the Sommerfeld expansions
yield\begin{equation}
G=G_{0}\tau(E_{F}),\quad Q=-\frac{k_{B}}{e}\frac{\pi^{2}}{3}\frac{\tau'(E_{F})}{\tau(E_{F})}k_{B}T,\label{eq:GandS_lowT}\end{equation}
where prime denotes a derivative. In our DFT-based results presented
below, we calculate $G$ and $Q$ according to Eq.~(\ref{eq:GandS}).
However, Eq.~(\ref{eq:GandS_lowT}) approximates the results to within
a few percent at room temperature, since $\tau(E)$ is smooth around
$E_{F}$ due to the off-resonant situation. It should be noted that
the equations neglect electron-vibration interactions. These could
in principle be included by adding the inelastic corrections to the
expression for the current,\cite{Viljas05} but we expect also these
contributions to be relatively small even at room temperature.

The transmission functions are computed with the help of Green's function
techniques. The electronic structure is described in terms of DFT
as implemented in the quantum chemistry program \textsc{\noun{Turbomole}}
\noun{V}5.7, where we employ the BP86 exchange-correlation functional
and the standard Gaussian basis set of split valence quality with
polarization functions on all non-hydrogen atoms.\cite{Ahlrichs89}
For further details on our method, see Refs.~\onlinecite{PaulyThesis,Pauly:NJP08,Wohlthat07,Pauly:PRB08}.

\section{Contacts\label{sec:Contacts}}

The molecules studied are shown in Fig.~\ref{cap:molecules}. Those
labeled with R1 to R4 are the pure oligophenylenes, S2 to S4 denote
oligophenylenes where the hydrogen atom in one of the two ortho positions
with respect to each ring-connecting carbon atom is substituted with
a methyl group,\cite{ElbingThesis,Shaporenko:2006} and D2 to D4 have
substituents in both ortho positions. Here, the numbers $N=1,\ldots,4$
refer to the number of phenyl rings in the molecule. The tilt angles
for the R, S, and D molecules vary between $33.4^{\circ}\leq\varphi_{R}\leq36.4^{\circ}$,
$84.8^{\circ}\leq\varphi_{S}\leq90^{\circ}$, and $88.8^{\circ}\leq\varphi_{D}\leq90^{\circ}$,
and the distances between the terminal carbon atoms of the molecules
are described to a good accuracy by $d=a+bN$, with $a=-0.154$ nm
and $b=0.435$ nm.

\begin{figure}
\begin{centering}\includegraphics[width=0.7\columnwidth,clip=]{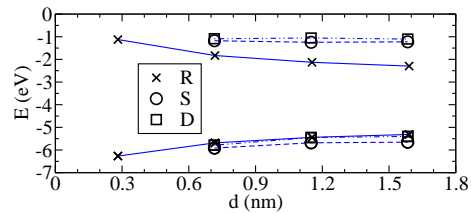}\par\end{centering}

\caption{\label{cap:HLplot}(Color online) HOMO and LUMO energies of the molecules
of Fig.~\ref{cap:molecules}.}
\end{figure}

The HOMO and LUMO energies of the molecules are shown in Fig.~\ref{cap:HLplot}.
It can be noticed that the HOMO-LUMO gaps of the S and D series are
larger than those of the R series.%
\begin{figure}[b]
\begin{centering}\includegraphics[width=0.9\columnwidth,clip=]{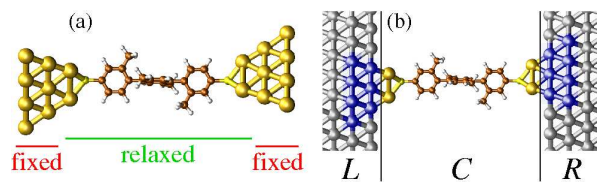}\par\end{centering}

\caption{\label{cap:junction}(Color online) (a) The contacts to the gold
electrodes are formed through sulfur atoms bonded to the hollow position
of the tip of a {[}111] pyramid. (b) For the transport calculations,
the tips of the pyramids are part of the extended molecule ($C$).
The more remote parts (blue) are absorbed into ideal semi-infinite
left ($L$) and ($R$) surfaces (gray).}
\end{figure}

To form the junctions, each molecule is coupled to the hollow position
of the tips of two gold {[}111] pyramids via a sulfur atom. The atomic
positions of the molecule and the first gold layers of the tips are
then relaxed. This is depicted in Fig.~\ref{cap:junction} for S3.
The relaxed part is also the {}``central region'' in the transport
calculations.\cite{Pauly:PRB08} The tilt angles $\varphi$ and the
distances $d$ of the contacted molecules are not essentially different
from those of the isolated molecules.

\section{Density-functional-based transport\label{sec:Density-functional-based-transport}}

\begin{figure}
\begin{centering}\includegraphics[width=1\columnwidth,clip=]{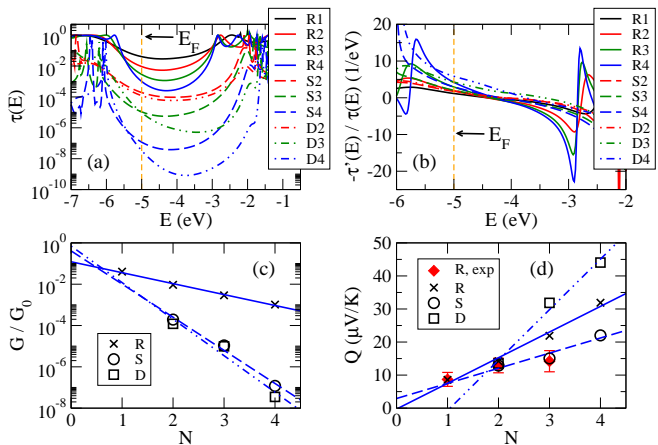}\par\end{centering}

\caption{\label{cap:DFTresults}(Color online) (a,b) Transmission function
and the negative of its logarithmic derivative. (c,d) The corresponding
$G$ and $Q$, including the temperature corrections in Eq.~(\ref{eq:GandS}).
The experimental data in (d) are from Ref.~\onlinecite{Reddy07}.
The straight lines are best fits to the numerical results for R (solid
line), S (dashed line), and D (dash-dot-dotted line).}
\end{figure}

The results for the transmission, its logarithmic derivative, the
conductance, and the thermopower for all of the R, S, and D type molecular
junctions are collected in Fig.~\ref{cap:DFTresults}. In order to
compare with the room-temperature experiments\cite{Reddy07} we set
$T=298$~K in Eq.~(\ref{eq:GandS}). The Seebeck coefficients are
displayed in Fig.~\ref{cap:DFTresults}(d) together with the experimental
results, where the molecules R1, R2, and R3 were studied.%

With increasing number $N$ of phenyl rings, the conductance decays
as $G/G_{0}\sim e^{-\beta N}$ {[}Fig.~\ref{cap:DFTresults}(c)].
For the R series, we find the decay coefficient $\beta_{R}=1.22$,\cite{Viljas07b}
in good agreement with theory of Ref.~\onlinecite{Kondo04}.\cite{FootnoteBetaR1}
Previous theoretical estimates\cite{Kaun03,Su05} and experimental
results\cite{Wold02,Wakamatsu06} for thiolated self-assembled monolayers
are consistently somewhat larger than this value. In particular, the
results reported in Ref.~\onlinecite{Wold02} vary between $1.5\leq\beta_{R}\leq2.1$,
and for amine end groups an experimental value of $\beta_{R}=1.5$
was reported.\cite{Venkataraman06} The underestimation of $\beta_{R}$
may indicate an overestimation of the conductance computed within
DFT.\cite{ToherPRL05,Quek07} However, a comparison between theory
and experiment is complicated due to the differences in the end groups
used.\cite{FootnoteBetaR2} Therefore, no conclusive comparison is
possible. For the S and D series, we find $\beta_{S}=3.69$ and $\beta_{D}=4.07$,
which are both much larger than $\beta_{R}$. This increase reflects
the reduced delocalization of the $\pi$-electron system. The absolute
conductance values of S and D are very similar to each other.

Since the Fermi energy at $E_{F}=-5.0$ eV lies closer to the HOMO
than to the LUMO level, the Seebeck coefficient $Q$ has a positive
value {[}Fig.~\ref{cap:DFTresults}(d)]. We also find that $Q$ increases
roughly linearly with $N$, as suggested by the experiments.\cite{Reddy07}
Indeed, assuming that the transmission around $E=E_{F}$ is of the
form $\tau(E)=\alpha(E)e^{-\beta(E)N}$, then Eq.~(\ref{eq:GandS_lowT})
yields $Q=Q^{(0)}+Q^{(1)}N$, where $Q^{(0)}=-k_{B}^{2}T\pi^{2}[\ln\alpha(E_{F})]'/3e$
and $Q^{(1)}=k_{B}^{2}T\pi^{2}\beta'(E_{F})/3e$.\cite{Viljas08}
Two things should be noted here. First, $Q$ does not necessarily
extrapolate to zero for $N=0$, leading to a finite {}``contact thermopower''
$Q^{(0)}$. Second, $Q^{(0)}$ depends on the prefactor $\alpha(E)$,
but $Q^{(1)}$ does not. Since $\alpha(E)$ contains the most significant
uncertainties related to the contact geometries, $Q^{(1)}$ can be
expected to be described at a higher level of confidence than $Q^{(0)}$.
Best fits to our results and the experimental data give $Q_{R}^{(0)}=-0.28$,
$Q_{R}^{(1)}=7.77$ and $Q_{R,exp}^{(0)}=6.43$, $Q_{R,exp}^{(1)}=2.75$
$\mu$V/K, respectively. Differences in the fit parameters mainly
stem from the data point for R3, where the experimental value is lower
than calculated. Considering the reported order-of-magnitude discrepancies
between measured conductances and those computed from DFT,\cite{ToherPRL05,Quek07}
the agreement still appears reasonable.\cite{Footnote3} However,
for large enough $T$ and $N$ the abovementioned exponential and
linear laws for the length dependences of $G$ and $Q$ should be
modified due to interactions with the thermal environment.\cite{Segal05}
The fact that the experimental data in Fig.~\ref{cap:DFTresults}(d)
exhibit a rather good linearity suggests that molecular vibrations
do not play a crucial role and that the electronic contribution to
$Q$ is dominant.%
\begin{figure}
\begin{centering}\includegraphics[width=0.9\columnwidth,clip=]{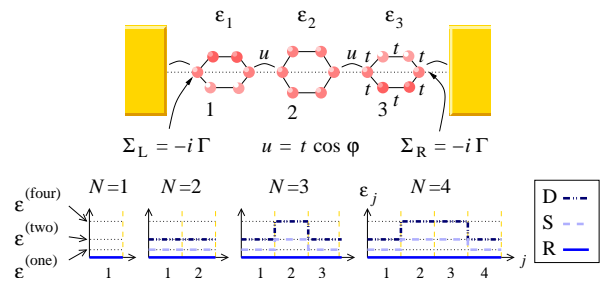}\par\end{centering}

\caption{\label{cap:model}(Color online) The parameters of the TB model represented
by a molecule with $N=3$ phenyl rings. Also shown are schematic graphs
of the onsite energies $\epsilon_{j}$ of the different rings $j=1,\ldots,N$
for R, S, and D type molecules with $N=1,\ldots,4$ rings.}
\end{figure}

Although the conductances for S and D are very similar, their thermopowers
are rather different. Furthermore, the magnitudes for $N>2$ follow
the surprising order $Q_{S}<Q_{R}<Q_{D}$.\cite{Footnote2} As we
will discuss below, these observations can be understood through the
two competing substituent effects: (i) a change in the alignment of
the $\pi$-electron levels with respect to $E_{F}$ as a result of
the electron-donating nature of the methyl group\cite{Venkataraman07,Baheti08}
and (ii) an increase in the ring-tilt angles. For the isolated molecules
the second effect results in the opening of the HOMO-LUMO gap when
going from molecules R to S or D, while the first effect causes the
difference between the HOMO and LUMO energies of the S and D series
(Fig.~\ref{cap:HLplot}).

\section{$\pi$-orbital model\label{sec:pi-orbital-model}}

In order to gain some understanding of the general features of the
dependence of $G$ and $Q$ on the number of substituents, we study
a simple tight-binding (TB) model, which describes the $\pi$-electron
system of the oligophenylenes (Fig.~\ref{cap:model}). The onsite
energies $\epsilon_{j}$ ($j=1,\ldots,N$) are equal on all carbon
atoms of phenyl ring $j$, the intra-ring hopping $t$ is assumed
to be the same everywhere, and the inter-ring hopping $u$ is parametrized
through $u=t\cos\varphi$. We assume the effect of the side groups
to come into play only through $\varphi$ and $\epsilon_{j}$. The
leads are modeled by ''wide-band'' self-energies $\Sigma_{L,R}=-i\Gamma$,
acting on the terminal carbon atoms.

\begin{figure}
\begin{centering}\includegraphics[width=1\columnwidth,clip=]{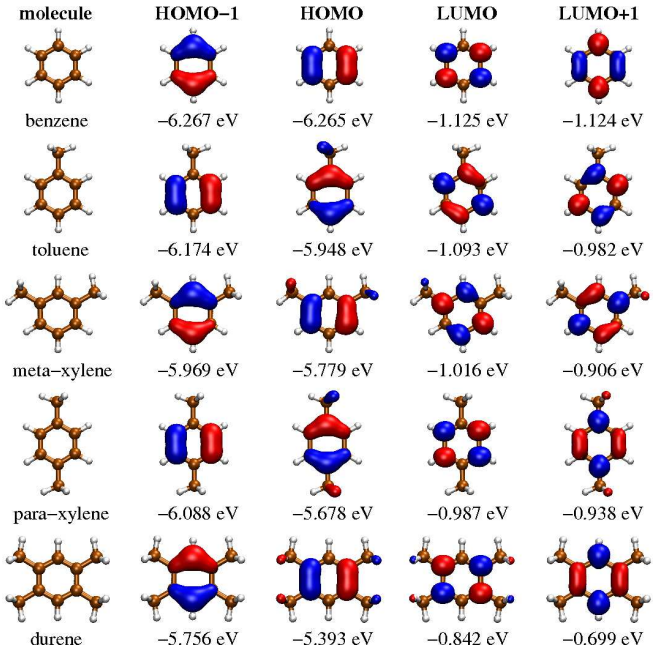}\par\end{centering}

\caption{\label{fig:DFT-benzene-shift}(Color online) Influence of methyl
substituents on the level alignment of frontier molecular orbitals
in benzene. From left to right the molecular structure is displayed
together with isosurface plots of the HOMO-1, HOMO, LUMO, and LUMO+1
wave functions. Below these plots the common name of the molecules
and the energies of the molecular levels are indicated.}
\end{figure}

\begin{figure}
\begin{centering}\includegraphics[width=1\columnwidth,clip=]{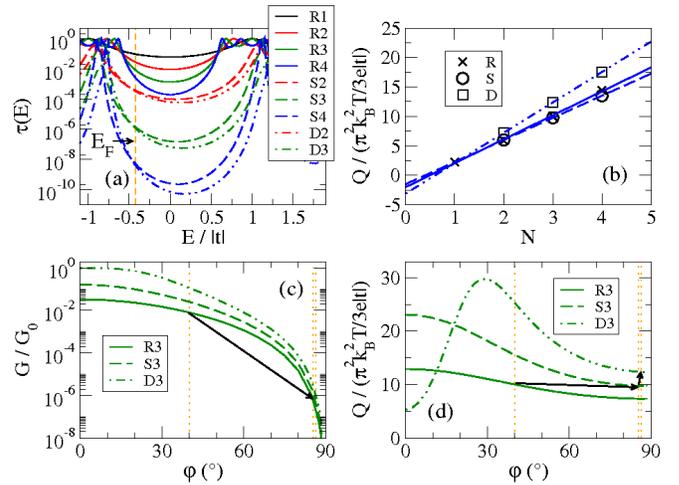}\par\end{centering}

\caption{\label{cap:modelresults}(Color online) (a) Transmission functions
for the TB model of Fig.~\ref{cap:model} with parameters chosen
as explained in the text. (b) Corresponding Seebeck coefficients according
to Eq.~(\ref{eq:GandS_lowT}). The straight lines are best fits as
in Fig.~\ref{cap:DFTresults}. (c,d) $G$ and $Q$ as a function
of the tilt angle for the molecules with $N=3$. The vertical dotted
lines indicate the {}``equilibrium angles'' $\varphi_{R}$, $\varphi_{S}$,
and $\varphi_{D}$, and the arrows represent changes when going from
R3 to S3 to D3.}
\end{figure}

We extract the parameters of our model as follows. For the R molecules,
we set $\epsilon_{j}=0$. In the S and D molecules, the $\epsilon_{j}$'s
for rings with one, two, or four methyl groups are obtained by performing
DFT calculations of methylbenzene, dimethylbenzene, and tetramethylbenzene
(Fig.~\ref{fig:DFT-benzene-shift}).%
 We find that the HOMO and LUMO energies in these molecules shift
upwards monotonously with the number of methyl substituents. This
can be attributed to the electron-donating character of the methyl
groups, and the resulting increase of Coulomb repulsion on the phenyl
ring. We use the averages of the HOMO and LUMO shifts relative to
benzene, and obtain, respectively, $\epsilon^{(one)}=0.17$, $\epsilon^{(two)}=0.33$,
and $\epsilon^{(four)}=0.58$ eV.\cite{Footnote1} The onsite energies
of the different phenyl rings of the R, S, and D molecules with $N=1,\ldots,4$
are schematically represented in the lower part of Fig.~\ref{cap:model}.
The hopping $t$ is set to half of (the negative of) the HOMO-LUMO
gap of benzene, with the result $t=-2.57$ eV (Fig.~\ref{cap:HLplot}).
The scattering rate $\Gamma$ and the tilt angles for R, S, and D
molecules are chosen to reproduce approximately the minimal transmissions
for R1, R2, S2, and D2 in the DFT results of Fig.~\ref{cap:DFTresults}(a).
This yields $\Gamma=0.64$ eV, $\varphi_{R}=40.0^{\circ}$, $\varphi_{S}=85.5^{\circ}$,
and $\varphi_{D}=86.5^{\circ}$. The last free parameter is $E_{F}$,
which should be determined by the overall charge-transfer effects
between the molecule and the electrodes. We choose its value of $E_{F}=-1.08$
eV close to the crossing points of the transmission curves of the
S and D molecules {[}cf.~Figs.~\ref{cap:DFTresults}(a) and \ref{cap:modelresults}(a)].%

Similar features can be recognized between the DFT results and our
model. In particular, as shown in Fig.~\ref{cap:modelresults}(b),
the correct order of the thermopowers is reproduced. In going from
R to S, the increase in the tilt angle and the associated breaking
of the $\pi$-electron conjugation dominates over other side group
effects. As a result both $G$ {[}Fig.~\ref{cap:modelresults}(c)]
and $Q$ {[}Fig.~\ref{cap:modelresults}(d)] decrease. In going from
S to D, an interplay of the sidegroup-induced level shifts and a small
residual increase in $\varphi$ raises $Q$ above the value for R,
while $G$ remains almost unchanged. We note that for an $N$-ring
junction the lowest-order terms in an expansion of the transmission
in powers of $u$ yield $\varphi$-dependences of the form $\tau(E)\approx c_{1}(E)\cos^{2(N-1)}\varphi+c_{2}(E)\cos^{2N}\varphi$,
for $E\approx E_{F}$.\cite{Pauly:PRB08} Independently of $N$, this
results in $Q\approx q_{1}+q_{2}\cos^{2}\varphi$, where we find $q_{1},q_{2}>0$
in our case. For $N=2$, the neglect of the $c_{2}$-term leads to
the well-known ``$\cos^{2}\varphi$ law'' of $G$,\cite{Venkataraman06,Pauly:PRB08}
but we see that for the tilt-angle dependence of $Q$ the presence
of this term is significant, since $q_{2}=0$ if $c_{2}=0$. When
$E_{F}$ is close to a resonance, further higher-order terms become
increasingly important and deviations from the $q_{1}+q_{2}\cos^{2}\varphi$
law result. This is seen most clearly as the non-monotonous $\varphi$-dependence
of $Q$ for molecule D3 in Fig.~\ref{cap:modelresults}(d).

\section{Effects beyond the $\pi$-orbital model\label{sec:beyond-pi-model}}

\begin{figure}
\begin{centering}\includegraphics[width=1\columnwidth,clip=]{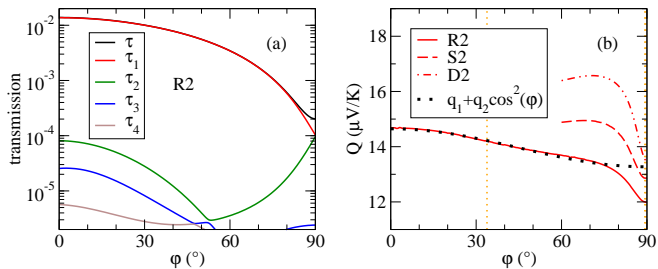}\par\end{centering}

\caption{\label{fig:TphiR2_SphiR2S2D2}(Color online) (a) Tilt-angle-dependent
transmission $\tau=\sum_{i}\tau_{i}$ resolved in its transmission
channels $\tau_{i}$ for the molecule R2 and (b) thermopower for R2,
S2, D2. The curve for R2 is fitted by a function of the form $q_{1}+q_{2}\cos^{2}\varphi$
with $q_{1}=13.27$ and $q_{2}=1.38$ $\mu$V/K for the tilt-angle
interval from $0^{\circ}$ to $60^{\circ}$. Dotted vertical lines
indicate DFT equilibrium tilt-angles $\varphi_{R2}$, $\varphi_{S2}$,
and $\varphi_{D2}$.}
\end{figure}

Close to perpendicular ring tilts results from the $\pi$-orbital
model should be taken with care. At $\varphi=90^{\circ}$ the $\pi$-$\pi$
coupling $u$ vanishes, and any other couplings between the rings
will become important.\cite{Woitellier:ChemPhys1989} Let us analyze
this for the biphenyl molecules R2, S2, and D2. For them, we have
varied the tilt angle between the rings, and have determined the charge
transport properties for every $\varphi$ using our DFT-based approach.
Further details of the procedure are described in Ref.~\onlinecite{Pauly:PRB08}.
For R2 (and similarly for S2 and D2) we observe that for most tilt
angles ($\varphi\lesssim80^{\circ}$) the transmission is dominated
by a single channel {[}Fig.~\ref{fig:TphiR2_SphiR2S2D2}(a)] of $\pi$-$\pi$
character.%
 However, at large angles two transmission channels of the same magnitude
are observed, which become degenerate at $\varphi=90^{\circ}$. They
arise from $\sigma$-$\pi$ couplings between the two rings, whose
strengths are proportional to $\sin\varphi$. Thus the two degenerate
channels are of the $\sigma$-$\pi$ and $\pi$-$\sigma$ type. These
features are obviously not accounted for by the TB model, where only
a single $\pi$-$\pi$ channel is present independently of $\varphi$.
The $\sigma$-$\pi$ couplings should also modify the tilt-angle dependence
of the thermopower, which is plotted in Fig.~\ref{fig:TphiR2_SphiR2S2D2}(b).
Due to the electron-donating nature of the methyl groups the thermopower
at fixed $\varphi$ increases from R2 to S2 and D2. While the curve
for R2 can be described by the law $q_{1}+q_{2}\cos^{2}\varphi$ for
$\varphi\lesssim80^{\circ}$, a dip is observed for larger $\varphi$.
Similar deviations in $Q$ are also present for S2 and D2, where we
have investigated a smaller tilt-angle interval because of the steric
repulsion of the methyl groups.\cite{Pauly:PRB08}

\begin{figure}
\begin{centering}\includegraphics[width=0.6\columnwidth,,clip=]{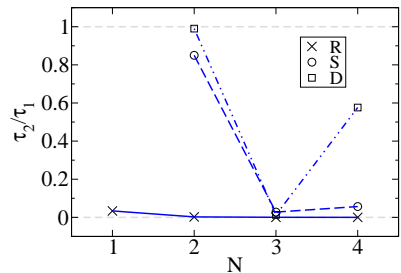}\par\end{centering}

\caption{\label{fig:t1overt2}(Color online) Importance $\tau_{1}/\tau_{2}$
of the second transmission channel $\tau_{2}$ as compared to the
first $\tau_{1}$ as a function of molecular length for the molecules
of series R, S, and D.}
\end{figure}

The degeneracy of the transmission channels is due to the $D_{2d}$
symmetry of biphenyl when $\varphi=90^{\circ}$.\cite{Woitellier:ChemPhys1989}
For the longer oligophenylenes the symmetry $D_{2d}$ can occur if
and only if tilt angles are all at $90^{\circ}$ and the number of
rings is even. Hence for S2, S4, D2, and D4 the ratio $\tau_{2}/\tau_{1}$
of the first two transmission channels should be particularly large.
Such even-odd oscillations are indeed visible in Fig.~\ref{fig:t1overt2}
for the S and D series, while they are absent for R.%
 Owing to the fact that tilt-angles deviate from $90^{\circ}$, the
oscillations decay. In particular $\tau_{2}/\tau_{1}$ is much smaller
for S4 than for D4, where the minimal tilt angles in the contacts
are $85.1^{\circ}$ and $89.1^{\circ}$, respectively.

\section{Discussion and conclusions\label{sec:Discussion-and-conclusions}}

As is well known, there are many theoretical uncertainties involved
in the determination of transport properties based on DFT calculations,
and improvements for the methods of molecular-scale transport theory
are currently being sought.\cite{ToherPRL05,Quek07,ThygesenPRL08}
Indeed, it is typical to find order-of-magnitude differences between
measured conductances and those computed within DFT.\cite{ToherPRL05,Quek07}
It is also known that atomic configurations can have a strong influence
on the conductance,\cite{XuePRB03,Stadler07,Quek07} and hence a conclusive
comparison with experimental data would require the statistical analysis
of a large number of contact geometries. However, models predict $Q$
to be insensitive to changes in the lead couplings,\cite{Paulsson03}
and hence it is expected to be a more robust quantity that $G$. Although
the measurements of Ref.~\onlinecite{Reddy07} were carried out at
room temperature, we do not expect molecular vibrations to play an
essential role in the results due to the weakness of the electron-vibration
coupling. Thus, the comparison we have made with our elastic transport
theory seems justified. Nevertheless, the investigation of the effect
of molecular vibrations is an interesting direction for future research,
also in view of optimizing the properties of molecular thermoelectric
devices for potential applications. As far as gaining a basic understanding
is concerned, low-temperature measurements would be desirable to remove
uncertainties related to the thermal excitation of vibrations. The
purpose of the present work was to study the general trends for a
series of molecules that are similarly coupled to the electrodes,
and considering all the potential uncertainties, the agreement we
obtain for the thermopower appears to be quite reasonable.

In conclusion, we have analyzed the length dependence of conductance
and thermopower for oligophenylene single-molecule contacts. While
we found the conductance to decay exponentially with the length of
the molecule, we observe that the thermopower increases linearly.
For possible future applications it is interesting to know, how the
magnitudes of these quantities can be tuned. We analyzed how this
can be achieved by chemical substituents, which control the Fermi-level
alignment and the molecular conformation. We demonstrated that a simple
$\pi$-electron tight-binding model can help to understand basic substituent
effects. In addition, we observed an even-odd effect for transmission
channel degeneracies upon variation of the number of phenyl rings,
which we explained by molecular symmetries.

\section{Acknowledgments\label{sec:Acknowledgments}}

We acknowledge stimulating discussions with M.~Mayor. The Quantum
Chemistry group of R.~Ahlrichs is thanked for providing us with \textsc{Turbomole}.
This work was financially supported by the Helmholtz Gemeinschaft
(Contract No.~VH-NG-029), the EU network BIMORE (Grant No. MRTN-CT-2006-035859),
and DFG SPP 1243. F.~P.~acknowledges the funding of a Young Investigator
Group at KIT.


\begin{thebibliography}{43}
\bibitem[{Foo({\natexlab{a}})}]{FootnoteBetaR1}
\bibinfo{note}{The decay coefficient reported in Ref.~\onlinecite{Kondo04} for
  the R series is $2.56$ nm$^{-1}$ as compared to our value of $\beta_R/b=2.80$
  nm$^{-1}$.}
\bibitem[{Foo({\natexlab{c}})}]{Footnote3}
\bibinfo{note}{We have performed similar calculations on another type of
  geometry, where the molecules are bonded to the gold electrodes in a "top"
  position.\cite{PaulyThesis} In this geometry the decay coefficients
  $\beta_{R,S,D}$ are by and large the same as presented above, but the
  thermopowers are higher by a factor of two.}
\bibitem[{Foo({\natexlab{d}})}]{Footnote2}
\bibinfo{note}{We obtain $Q^{(0)}_{S}=2.93$, $Q^{(1)}_{S}=4.57$ and
  $Q^{(0)}_{D}=-16.03$, $Q^{(1)}_{D}=15.28$ $\mu$V/K from best fits.}
\bibitem[{Foo({\natexlab{e}})}]{Footnote1}
\bibinfo{note}{For dimethylbenzene we average over the "meta" and "para"
  types.}
\end{thebibliography}
\end{document}